\newcommand{\mystyleflag}{0}

\ifnum\mystyleflag=0
\documentclass[12pt]{article}
\pdfoutput=1
\usepackage{jcappub}
\else
\documentclass[journal,article,submit,pdftex,moreauthors]{Definitions/mdpi} 
\firstpage{1} 
\pubvolume{1}
\issuenum{1}
\articlenumber{0}
\pubyear{2024}
\copyrightyear{2024}
\datereceived{ } 
\daterevised{ } 
\dateaccepted{ } 
\datepublished{ } 
\fi

\usepackage{amsmath}
\usepackage{amssymb}
\usepackage{physics}
\usepackage{slashed}
\usepackage{subfigure}
\usepackage{float}
\usepackage{notoccite}
\usepackage{cancel}
\def\be{\begin{equation}}
\def\ee{\end{equation}}
\def\bea{\begin{eqnarray}}
\def\eea{\end{eqnarray}}

\ifnum\mystyleflag=0
\title{Emergent field theories from neural networks}
\author[1,2]{Vitaly Vanchurin} 
\emailAdd{vitaly.vanchurin@gmail.com}
\affiliation[1]{Artificial Neural Computing, Weston, Florida, 33332, USA}
\affiliation[2]{Duluth Institute for Advanced Study, Duluth, Minnesota, 55804, USA}
\begin{document}   
\else
\Title{Emergent field theories from neural networks}
\Author{Vitaly  Vanchurin $^{1,2}$}
\AuthorNames{Vitaly  Vanchurin}
\AuthorCitation{Vanchurin, V.}
\address{%
$^{1}$ \quad Artificial Neural Computing, Weston, Florida, 33332, USA\\
$^{2}$ \quad Duluth Institute for Advanced Study, Duluth, Minnesota, 55804, USA}
\fi

\abstract{We establish a duality relation between Hamiltonian systems and neural network-based learning systems. We show that the Hamilton's equations for position and momentum variables correspond to the equations governing the activation dynamics of non-trainable variables and the learning dynamics of trainable variables. The duality is then applied to model various field theories using the activation and learning dynamics of neural networks. For Klein-Gordon fields, the corresponding weight tensor is symmetric, while for Dirac fields, the weight tensor must contain an anti-symmetric tensor factor. The dynamical components of the weight and bias tensors correspond, respectively, to the temporal and spatial components of the gauge field. }

\ifnum\mystyleflag=0
\maketitle  
\else
\keyword{Neural Networks, Emergent fields, Klein-Gordon field, Dirac Field,  Gauge fields} 
\begin{document}   
\fi

\section{Introduction}

An artificial neural network can be understood as a complex dynamical system exhibiting three distinct types of dynamics. First, boundary dynamics, which involves updating the input and output neurons according to data from the training dataset. Second, activation dynamics, where neurons process and transmit information through connections within the network. Finally, learning dynamics governs the adaptation of the network, where biases and connection weights between neurons are adjusted in order to reduce a loss function, ultimately enhancing the network's ability to accurately model the training data. See Refs \cite{Galushkin, Haykin, TTML} for mathematical introductions to the subject of artificial neural network.

Similarly, any physical system, such as one described by classical field theory, can be viewed as a dynamical system, but with only two types of dynamics, which we refer to as bulk and boundary. In particular, equations of motion, such as those in field theories, must be defined within the bulk, and boundary conditions must be specified whenever necessary. Therefore, to establish an equivalence between the dynamical systems of neural networks and field theories, we need to understand how the bulk dynamics can emerge from the activation and learning dynamics within a neural network. This goes beyond the straightforward idea that a sufficiently complex neural network could learn the bulk dynamics of any physical system. We are not interested in supervised learning of the bulk dynamics; instead, we aim to determine when the bulk dynamics could emerge from the activation and learning dynamics, or when the equations of motion of a given field theory could emerge from the equations governing activation and learning. 

The paper is organized as follows. In Sec. \ref{Sec:Hamiltonian}, we establish a duality relation between Hamiltonian systems and neural network-based learning systems. In Sec. \ref{Sec:Klein-Gordon}, the duality is applied to demonstrate how Klein-Gordon fields can emerge from the activation and learning.  In Sec. \ref{Sec:Dirac}, we demonstrate that the Dirac equation can also emerge from a learning system if the weight tensor includes an anti-symmetric factor that describes the interaction between different types of neurons. In Sec. \ref{Sec:Gauge}, we show that the Dirac equation becomes covariant with respect to gauge transformations if the weight and bias tensors contain dynamical components. Finally, in Sec. \ref{Sec:Discussion}, we summarize and discuss the main results of the paper.

\section{Hamiltonian system}\label{Sec:Hamiltonian}

Consider a single neuron described by a non-trainable state $\phi$, a trainable bias $\beta$, and a fixed self-connection weight $W = 1$. Then, the activation dynamics for a linear activation function is given by
\be
\phi(t+1) = \beta(t) + W \phi(t) = \beta(t) + \phi(t) \label{eq:discrete1}
\ee
where $\phi(t+1)$ represents the state at the next time step. The learning dynamics, using gradient descent, is given by
\be
\beta(t+1) = \beta(t) - \frac{d H(\phi,\beta)}{d \beta}. \label{eq:discrete2}
\ee
where $H(\phi, \beta)$ is the loss function.We assume a mini-batch size of one and that only a single step of back-propagation is applied, i.e.
\be
\frac{\partial \phi}{\partial \beta} = 1.
\ee
Therefore the update rule for $\beta$ becomes
\be
\beta(t+1) = \beta(t) - \frac{\partial H(\phi, \beta)}{\partial \beta} - \frac{\partial H(\phi, \beta)}{\partial \phi}. \label{eq:discrete3}
\ee
where the gradient of loss function $H(\beta, \phi(\beta))$ with respect to $\beta$ includes both the explicit dependence of $H$ on $\beta$ and the implicit dependence through $\phi(\beta)$.

In a continuous-time limit equations \eqref{eq:discrete1} and \eqref{eq:discrete2} can be expressed as
\bea
\frac{d\phi}{d t} &=&  \beta  \label{eq:continuum}\\
\frac{d \beta}{d t} &=& - \frac{\partial H(\phi, \beta)}{\partial \beta} - \frac{\partial H(\phi, \beta)}{\partial \phi} \notag
\eea
which are to be compared with the Hamilton's equations:
\bea
\frac{d\phi}{d t} &=& \frac{\partial {\cal H}(\phi,\beta)}{\partial \beta}   \label{eq:hamiltonian}\\
\frac{d \beta}{d t} &=& - \frac{\partial {\cal H}(\phi,\beta)}{\partial \phi} \notag
\eea
where ${\cal H}(\phi, \beta)$ is the Hamiltonian. Evidently, for \eqref{eq:continuum} and \eqref{eq:hamiltonian} to coincide, the Hamiltonian must be given by 
\be
{\cal H}(\phi, \beta) = \frac{1}{2} \beta^2 + {\cal V}(\phi)
\ee
where
\be
\frac{d {\cal V}(\phi)}{d \phi}    =   \frac{\partial H(\phi, \beta)}{\partial \beta} + \frac{\partial H(\phi, \beta)}{\partial \phi}.
\ee
By solving it for the loss function $H(\phi, \beta)$ we obtain
\be
H(\phi, \beta) = {\cal V}(\phi)  + F(\phi - \beta) 
\ee
where $F(\cdot)$ is an arbitrary function that may include a constant.

Next, consider $N$ neurons described by a state vector $\phi$,  a fixed weight matrix  $W$, and a bias vector $\beta$, which is a linear function of trainable variables, 
\be
\beta_i = B_i^{\phantom{i}j} \pi_j,
\ee
where Einstein summation convention over repeated indices is implied throughout the paper unless stated otherwise. For simplicity, we also assume that the bias matrix $B$ is invertible. Then, the activation and learning dynamics in the continuous-time limit are given by
\bea
\frac{d\phi_i}{d t} &=&  (W-I)_i^{\phantom{i}j} \phi_j +  B_i^{\phantom{i}j} \pi_j \\
\frac{d \pi_i}{d t} &=& - \delta_{ij}  \frac{d H(\phi,\pi)}{d \pi_j} = -\delta_{ij} \frac{\partial H(\phi,\pi)}{\partial \pi_j} - (B^T)_{ij}  \frac{\partial H(\phi,\pi)}{\partial \phi_j}  \notag
\eea
or 
\bea
\frac{d\phi_i}{d t} &=& (W-I)_i^{\phantom{i}j} \phi_j +  \beta_i  \label{eq:continuum2}\\
 \frac{d \beta_i}{d t} &=& - \left (BB^T \right )_{ij}   \left ( \frac{\partial H(\phi,\pi)}{\partial \beta_j}+ \frac{\partial H(\phi,\pi)}{\partial \phi_j} \right ) \notag
\eea
where $\delta_{ij}$ is the Kronecker delta. These equations would be equivalent to the Hamilton's equations: 
\bea
\frac{d\phi_i}{d t} &=& \delta_{ij} \frac{\partial {\cal H}(\phi,\beta)}{\partial \beta_j}   \label{eq:hamiltonian2}\\
\frac{d \beta_i}{d t} &=& - \delta_{ij} \frac{\partial {\cal H}(\phi,\beta)}{\partial \phi_j} \notag,
\eea
 if the Hamiltonian function is given by 
\be
{\cal H}(\phi, \beta) = \frac{1}{2} \beta^2 + \beta^i (W-I)_i^{\phantom{i}j} \phi_j + {\cal V}(\phi) \label{eq:hamilton}
\ee
where 
\be
((W-I)^T)_i^{\phantom{i}j}  \beta_j +\frac{\partial {\cal V}(\phi)}{\partial \phi_i}    = \left (BB^T \right )_{ij}   \left ( \frac{\partial H(\phi,\pi)}{\partial \beta_j}+ \frac{\partial H(\phi,\pi)}{\partial \phi_j} \right ).
\ee
Under assumption that the weight matrix is symmetric, i.e. $W=W^T$, and the bias and weight matrices commute, i.e. $W B = B W$, we obtain a solution for the loss function 
\be
H(\phi, \pi) =  \frac{1}{2} \pi^i (W-I)_i^{\phantom{i}j}  \pi_j  + \sum_{i,j}   \left ( \int  d\phi_i \left (B^{-T} B^{-1} \right )_{ij} \frac{\partial {\cal V}\left (\phi\right )}{\partial \phi_j}\right )  + \sum_i F_i(\phi_i - \sum_j B_i^{\phantom{i}j} \pi_j)\label{eq:loss}
\ee
where $F_i(\cdot)$ are arbitrary functions that may include constants. Therefore, any physical system that can be described by the Hamiltonian function \eqref{eq:hamilton} has a dual representation as a learning system described by the loss function \eqref{eq:loss}, assuming a symmetric weight matrix $W$ and invertible bias matrix $B$. 

In general, any learning system can be described to linear order by
\bea
\frac{d\phi_i}{d t} &=&  (W-I)_i^{\phantom{i}j} \phi_j +  B_i^{\phantom{i}j} \pi_j \\
\frac{d \pi_i}{d t} &=&  A_i^{\phantom{i}j} \phi_j + (V-I)_i^{\phantom{i}j} \pi_j \notag
\eea
or, assuming $B$ is invertible, by
\bea
\frac{d\phi_i}{d t} &=&  (W-I)_i^{\phantom{i}j} \phi_j +  \beta_i \\
\frac{d \beta_i}{d t} &=&  (BA)_i^{\phantom{i}j} \phi_j + (B(V-I)B^{-1})_i^{\phantom{i}j} \beta_j. \notag
\eea
Moreover, if
\bea
V &=& 2 I - B^{-1} W B\notag\\
A &=& -B^T\\
W &=& W^T\notag,
\eea
then the learning system defined by the loss function
\be
H(\phi,\pi) = \frac{1}{2} \phi^i  \phi_i  -  \frac{1}{2} \pi^i \pi_i + \frac{1}{2} \pi^i ( B^{-1} W B)_i^j \pi_j
\ee
is dual to a physical system defined by the Hamiltonian function 
\be
{\cal H(\phi, \beta)} = \frac{1}{2} \phi^i (B B^T)_i^j \phi_j + \frac{1}{2} \beta^i \beta_i + \beta^i (W-I) \phi_i.
\ee
However, many learning phenomena are intrinsically non-linear and, therefore, would not be captured by this simple Hamiltonian description.
 
\section{Klein-Gordon field}\label{Sec:Klein-Gordon}

In the previous section, we established a duality between learning systems, represented by a loss function $H(\phi,\pi)$ and physical systems, represented by a Hamiltonian function ${\cal H}(\phi,\beta)$. To construct this duality, we introduce the bias matrix $B$, defining biases as a linear function of the trainable parameters, i.e. $\beta_i = B_i^{\phantom{i}j} \pi_j$, and assumed that the weight matrix $W$ is symmetric, i.e. $W=W^T$. In this section, we will apply the duality to describe the Klein-Gordon field as a learning system, and in the subsequent sections, we will extend this approach to the Dirac field by considering anti-symmetric tensor factors of the weight and bias tensors. 

Consider the Klein-Gordon field equation in $3+1$ dimensions:
\be
\eta_{\mu\nu}\frac{\partial }{\partial x_\mu}\frac{\partial }{\partial x_\nu}  \phi({x}) +  m^2 \phi({x}) = 0,
\ee
where $\eta_{\mu\nu}$ is the flat metric with signature $\left(+,-, -, -\right)$ and $\mu,\nu \in \{0,1,2,3\}$. The system 
can be described in terms of Hamilton's equations:
\bea
\frac{\partial \phi({x})}{\partial x_0} &=& \beta({x}) \\
\frac{\partial \beta({x})}{\partial x_0} &=&   \sum_{\mu=1}^3  \frac{\partial^2 \phi({ x})}{\partial x_\mu^2}  - m^2 \phi({x}) \notag
\eea
with the Hamiltonian 
\be
{\cal H}[\phi,\beta] = \int dx^D \left ( \frac{1}{2}  \beta({ x})^2 + \frac{1}{2} \sum_{\mu=1}^3 \left (\frac{\partial \phi({ x})}{\partial x_\mu}\right )^2 +\frac{1}{2} m^2 \phi({x})^2 \right ).
\ee
In the limit of discretized space (i.e., with lattice vectors ${\bf e}_1=(1,0,0)$, ${\bf e}_2=(0,1,0)$, ${\bf  e}_3=(0,0,1)$) and time (i.e., with unit time-step), the Hamilton's equations take the following form: 
\bea
\phi_{\bf i}(t+1) - \phi_{\bf i}(t)&=& \beta_{\bf i}(t) \\
\beta_{\bf i}(t+1) - \beta_{\bf i}(t) &=& \sum_{\mu =1}^3 \left (  \phi_{\bf i + e_\mu} (t) - 2\phi_{\bf i } (t) + \phi_{\bf i - e_\mu} (t)   \right ) - m^2 \phi_{\bf i}(t),\notag
\eea
where ${\bf i} = (i_1, i_2,i_3)$, for the Hamiltonian 
\be
{\cal H}(\phi,\beta)= \frac{1}{2}  \pi^{\bf i} \pi_{\bf i}-  \frac{1}{2} \sum_{\mu=1}^3 \phi^{\bf i}  \left (  \phi_{\bf i + e_\mu}  - 2\phi_{\bf i } + \phi_{\bf i - e_\mu}    \right )  +\frac{1}{2} m^2 \phi^{\bf i} \phi_{\bf i}.
\ee

Equivalently, the Klein-Gordon equation can be described as a learning system with non-trainable and trainable variables described as tensors:
\bea
\phi_{\bf{i}} &\equiv & \phi_{i_1,i_2,i_3} \\
\pi_{\bf{i}} &\equiv & \pi_{i_1,i_2,i_3}. \notag
\eea
and their dynamics given by
\bea
\phi_{\bf{i}}(t+1) &=& W^{\bf{j}}_{{\bf i}} \phi_{\bf{j}}(t) + B^{\bf{j}}_{{\bf i}}  \pi_{\bf{j}}(t) \\
\pi_{\bf{i}} (t+1) &=& A^{\bf{j}}_{{\bf i}} \phi_{\bf{j}}(t) + V^{\bf{j}}_{{\bf i}} \pi_{\bf{j}}(t) \notag
\eea
where
\bea
A_{\bf i}^{\phantom{i}\bf{j}} &=& -\sum_{\mu=1}^3 \left ( \delta_{\bf i}^{\phantom{i}\bf j - e_\mu} - 2\delta_{\bf i}^{\phantom{i}\bf j} + \delta_{\bf i}^{\phantom{i}\bf j + e_\mu} \right ) + m^2  \delta_{\bf i}^{\phantom{i}\bf j}\notag \\
B_{\bf i}^{\phantom{i}\bf{j}} &=& \delta_{\bf i}^{\phantom{i}\bf j}  \notag\\
W_{\bf i}^{\phantom{\bf i}\bf j}&=& \delta_{\bf i}^{\phantom{i} \bf j} \\
V_{\bf i}^{\phantom{i}\bf{j}} &=& \delta_{\bf i}^{\phantom{i}\bf j}. \notag 
\eea
This corresponds to the activation dynamics
\be 
\phi_{\bf i}(t+1) = \phi_{\bf i}(t) + \beta_{\bf i}(t)\label{eq:act1}
\ee
and the learning dynamics, using gradient decent, 
\be
\beta_{\bf i}(t+1) -  \beta_{\bf i}(t)  =  \sum_{\mu=1}^3 \left (  \phi_{\bf i + e_\mu}(t)  - 2\phi_{\bf i }(t) + \phi_{\bf i - e_\mu}(t)    \right )  -  m^2 \phi_{\bf i}(t)\label{eq:lrn1}
\ee
with the loss function 
\be
{H}(\phi,\beta)= -  \frac{1}{2}\sum_{\mu=1}^3 \phi^{\bf i}  \left (  \phi_{\bf i + e_\mu}  - 2\phi_{\bf i } + \phi_{\bf i - e_\mu}    \right )  +  \frac{1}{2} m^2 \phi^{\bf i} \phi_{\bf i}. 
\ee
As in the previous section, we used 
\be
\frac{\partial \phi_{\bf i}}{\partial \beta_{\bf j}} = \delta_{\bf i}^{\phantom{i}\bf j}
\ee
which follows from activation dynamics \eqref{eq:act1}. By combing equations \eqref{eq:act1} and \eqref{eq:lrn1}, we obtain a discrete version of the Klein-Gordon equation:
\be
\phi_{\bf i}(t+2) -  2\phi_{\bf i}(t+1) +\phi_{\bf i}(t) = \sum_{\mu=1}^3 \left (  \phi_{\bf i + e_\mu}(t)  - 2\phi_{\bf i }(t) + \phi_{\bf i - e_\mu}(t)    \right )  -  m^2 \phi_{\bf i}(t). \label{eq:KG1}
\ee

Note, however, that the discrete representation of the Klein-Gordon field is not unique, and there are other combinations of activation and learning dynamics that correspond to the same equations in the continuous limit. For example, consider
\bea
A_{\bf i}^{\phantom{i}\bf{j}} &=& -\delta_{\bf i}^{\phantom{i}\bf j}\notag \\
B_{\bf i}^{\phantom{i}\bf{j}} &=& \delta_{\bf i}^{\phantom{i}\bf j} \notag\\
W_{\bf i}^{\phantom{\bf i}\bf j}&=& -(4 +m^2) \delta_{\bf i}^{\phantom{i}\bf j} +\sum_{\mu=1}^3  \left (\delta_{\bf i}^{\phantom{i}\bf j-e_\mu} +\delta_{\bf i}^{\phantom{i}\bf  j+e_\mu}  \right)  \\
V_{\bf i}^{\phantom{i}\bf{j}} &=& 0\notag 
\eea
In this case, the activation dynamics is 
\be
\phi_{\bf i}(t+1)  =- (4+m^2)\phi_{\bf i}(t) +\sum_{\mu=1}^3 \left (  \phi_{\bf i + e_\mu}(t)  +\phi_{\bf i - e_\mu}(t)  \right ) +\beta_{\bf i}(t)
\label{eq:act2},
\ee 
and the learning dynamics, using gradient decent, is
\be
\beta_{\bf i}(t+1) -\beta_{\bf i}(t)  = - \beta_{\bf i}(t) - \phi_{\bf i}(t) 
\ee
or
\be
\beta_{\bf i}(t+1)  = - \phi_{\bf i}(t) \label{eq:lrn2}.
\ee
with the loss function 
\be
H(\phi,\beta) =  \frac{1}{2} \phi^{\bf i} \phi_{\bf i} +\frac{1}{2} \beta^{\bf i} \beta_{\bf i}.
\ee
By combing equations \eqref{eq:act2} and \eqref{eq:lrn2}, we also obtain a discrete version of the Klein-Gordon equation:
\be
\phi_{\bf i}(t+1) -  2\phi_{\bf i}(t) +\phi_{\bf i}(t-1) = \sum_{\mu=1}^3 \left (  \phi_{\bf i + e_\mu}(t)  - 2\phi_{\bf i }(t) + \phi_{\bf i - e_\mu}(t)    \right )  -  m^2 \phi_{\bf i}(t)\label{eq:KG2},
\ee
but the discrete second derivative is now evaluated between $t-1$ and $t+1$, as opposed to between $t$ and $t+2$ in equation \eqref{eq:KG1}. Therefore, the duality between learning systems and field theories is not a map, but a relation --- a many-to-one relation.

\section{Dirac field}\label{Sec:Dirac}

Consider $K$ distinct types of neurons, corresponding to the configuration space, on a $D$-dimensional lattice, corresponding to the physical space. The states of both non-trainable and trainable variables then can be described as tensors:
\bea
\phi^{a}_{\bf{i}} &\equiv & \phi^{a}_{i_1,...,i_D} \\
\pi^{a}_{\bf{i}} &\equiv & \pi^{a}_{i_1,...,i_D}, \notag
\eea
where $a \in \{1,...,K\}$, and their dynamics to linear order is given by:
\bea
\phi^{a}_{\bf{i}}(t+1) &=& W^{a, \bf{j}}_{{\bf i},b} \phi^{b}_{\bf{j}}(t) + B^{a, \bf{j}}_{{\bf i},b}  \pi^{b}_{\bf{j}}(t) \label{eq:eom}\\
\pi^{a}_{\bf{i}} (t+1) &=& A^{a, \bf{j}}_{{\bf i},b} \phi^{b}_{\bf{j}}(t) + V^{a, \bf{j}}_{{\bf i},b} \pi^{b}_{\bf{j}}(t). \notag
\eea
If the loss function is given by
\be
H(\phi,\pi) =  \frac{1}{2} \phi_a^{\bf i} \phi^a_{\bf i} +\frac{1}{2} \pi_a^{\bf i} \pi^a_{\bf i} - \frac{1}{2}   \pi_a^{\bf i}  V^{a, \bf{j}}_{{\bf i},b}  \pi^b_{\bf j}. 
\ee
then
\bea
A^{a, \bf{j}}_{{\bf i},b} \phi^{b}_{\bf{j}} &=& - \frac{\partial  \phi_{b}^{\bf{j}}}{\partial \pi_{a}^{\bf{i}}}\phi^{b}_{\bf{j}} =  -(B^T)^{a, \bf{j}}_{{\bf i},b}\phi^{b}_{\bf{j}}
\eea
which implies
\bea
A &=& -B^T.
\eea
In this case equation \eqref{eq:eom} becomes
\bea
\phi^{a}_{\bf{i}}(t+1) - \phi^{a}_{\bf{i}}(t)  &=& \left ( W- I\right )^{a, \bf{j}}_{{\bf i},b} \phi^{b}_{\bf{j}}(t)+B^{a, {\bf j}}_{b,{\bf i}} \pi^{b}_{\bf{j}}(t) \label{eq:eom2}\\
\pi^{a}_{\bf{i}}(t+1) - \pi^{a}_{\bf{i}}(t) &=& -(B^T)^{a, \bf{j}}_{{\bf i},b} \phi^{b}_{\bf{j}}(t)+ \left (V-I  \right)^{a, {\bf j}}_{b,{\bf i}} \pi^{b}_{\bf{j}}(t) \notag
\eea
and if $V$ and $B$ commute then
\bea
\phi^{a}_{\bf{i}}(t+1) - \phi^{a}_{\bf{i}}(t)   &=&  \left ( W- I\right )^{a, \bf{j}}_{{\bf i},b} \phi^{b}_{\bf{j}}(t)+\beta^{a}_{\bf{i}}(t) \label{eq:eom3}\\
\beta^{a}_{\bf{i}}(t+1) - \beta^{a}_{\bf{i}}(t)  &=& -\left ( B B^T \right) ^{a, \bf{j}}_{{\bf i},b} \phi^{b}_{\bf{j}}(t) + \left (V-I  \right)^{a, {\bf j}}_{b,{\bf i}} \beta^{b}_{\bf{j}}(t) \notag
\eea
where as before 
\be
\beta^{a}_{\bf{i}} \equiv B^{a, {\bf j}}_{b,{\bf i}}  \pi^{b}_{\bf{j}}
\ee
corresponds to biases. 

By substituting the first equation in \eqref{eq:eom3} into the second equation, we obtain: 
\bea
\phi^a_{\bf i}(t+2) -  2\phi^a_{\bf i}(t+1) +\phi^a_{\bf i}(t) &=& \left ( - B B^T +(V-I)(I-W) \right)^{a, \bf{j}}_{{\bf i},b} \phi^{b}_{\bf{j}}(t) +  \label{eq:KG3}\\
&+& (V+ W-2 I)^{a, \bf{j}}_{{\bf i},b}   (\phi^{b}_{\bf{j}}(t+1) - \phi^{b}_{\bf{j}}(t)). \notag
\eea
which simplifies for $V = 2 I - W$ 
\be
\phi^a_{\bf i}(t+2) -  2\phi^a_{\bf i}(t+1) +\phi^a_{\bf i}(t) = \left ( - B B^T - (W-I)^2 \right)^{a, \bf{j}}_{{\bf i},b} \phi^{b}_{\bf{j}}(t).\label{eq:KG4}
\ee
The corresponding equations of activation and learning dynamics are then given by:
\bea
\phi^{a}_{\bf{i}}(t+1) - \phi^{a}_{\bf{i}}(t)  &=& \left ( W- I\right )^{a, \bf{j}}_{{\bf i},b} \phi^{b}_{\bf{j}}(t)+B^{a, {\bf j}}_{b,{\bf i}} \pi^{b}_{\bf{j}}(t) \label{eq:eom4}\\
\pi^{a}_{\bf{i}}(t+1) - \pi^{a}_{\bf{i}}(t) &=& -(B^T)^{a, \bf{j}}_{{\bf i},b} \phi^{b}_{\bf{j}}(t)+ \left (I -W \right)^{a, {\bf j}}_{b,{\bf i}} \pi^{b}_{\bf{j}}(t). \notag
\eea
which can be realized, for example, for the loss function 
\be
H(\phi,\pi) =  \frac{1}{2} \phi_a^{\bf i} \phi^a_{\bf i} - \frac{1}{2}  \pi_a^{\bf i} \pi^a_{\bf i}  + \frac{1}{2}   \pi_a^{\bf i}  W^{a, \bf{j}}_{{\bf i},b}  \pi^b_{\bf j} 
\ee
where the weight matrix $W$ is symmetric, i.e. $W=W^T$.

More generally, the weight matrix in \eqref{eq:eom4} need not be symmetric. For example, consider a weight matrix which can be expressed as a tensor product
\be
(W-I)^{a, \bf{j}}_{{\bf i},b} = m {\cal X}^a_{0,b} \delta^{ \bf j}_{ \bf i}
\ee
where  ${\cal X}^a_{0,b}  = - {\cal X}^b_{0,a}$ is anti-symmetric factor. Then the equation for learning dynamics in \eqref{eq:eom4} is realized, for example, if different types of neurons are described by different loss functions, i.e.
\be
H_a(\phi,\pi) =  \frac{1}{2} \phi_a^{\bf i} \phi^a_{\bf i} -  m  \pi_a^{\bf i} {\cal X}^a_{0,b} \pi^b_{\bf i} \label{eq:losses}
\ee
with no summation over $a \in \{1,....,K\}$. In what follows, we will generalize the duality between physical and learning systems by introducing anti-symmetric tensor factors in the weight $W$ and bias $B$ tensors.

\subsection{Two types of neurons on one-dimensional lattice} 

For starters, consider $K=2$ types of neurons on $D=1$ dimensional lattice with 
\bea
(W-I)^{a, {j}}_{{i},b} =  m {\cal X}^a_{0,b} \delta^{ j}_{ i}  &=&  m \epsilon^a_b  \delta^{ j}_{ i} \label{eq:WB} \\
B^{a, j}_{i,b} =  {\cal X}^a_{1,b}  \left ( \delta^{j-1}_{i}  - \delta^{j}_i \right) &=& \epsilon^a_b  \left ( \delta^{j-1}_{i}  - \delta^{j}_i \right). \notag
\eea
where the anti-symmetric factors are given by the Levi-Civita symbol
\be
{\cal X}^a_{0,b} = {\cal X}^a_{1,b}  = \epsilon^a_b.
\ee
By substituting the weight and bias tensors \eqref{eq:WB} into \eqref{eq:eom5} we obtain:
 \bea
\phi^{1}_{i}(t+1) - \phi^{1}_{i}(t)  &=& ( \pi^{2}_{i+1}(t) - \pi^{2}_{i}(t) ) + m \phi^{2}_{i}(t) \notag\\
\pi^{1}_{i}(t+1) - \pi^{1}_{i}(t) &=& - (\phi^{2}_{{i}}(t) - \phi^{2}_{{i-1}}(t))  - m \pi^{2}_{i}(t)\\
\phi^{2}_{i}(t+1) - \phi^{2}_{i}(t)  &=& - (\pi^{1}_{i+1}(t) - \pi^{1}_{i}(t)) - m \phi^{1}_{i}(t)\notag\\
\pi^{2}_{i}(t+1) - \pi^{2}_{i}(t) &=& (\phi^{1}_{i}(t) - \phi^{1}_{i-1}(t)) + m \pi^{1}_{i}(t)\notag
\eea
and in the continuous limit, this leads to: 
\bea
\frac{\partial}{\partial x^0}\phi^1(x) - \frac{\partial}{\partial x^1}\pi^2(x) - m \phi^2(x) &=& 0 \notag\\
\frac{\partial}{\partial x^0} \pi^1(x)+ \frac{\partial}{\partial x^1} \phi^2(x) + m \pi^2(x)  &=&  0\label{eq:Dirac1}\\
\frac{\partial}{\partial x^0} \phi^2(x)+\frac{\partial}{\partial x^1} \pi^1(x)+ m \phi^1(x) &=&  0 \notag\\
\frac{\partial}{\partial x^0} \pi^2(x)-\frac{\partial}{\partial x^1}\phi^1(x) - m \pi^1(x)&=&  0\notag
\eea
for position $x^1=i$ and time $x^0=t$ coordinates. Furthermore, if we define a two-component spinor as
 \be
 \psi \equiv \begin{pmatrix}
 {\phi}^{1} + i {\phi}^{2}\\
 {\pi}^{1} + i {\pi}^{2}
 \end{pmatrix}
 \ee
then \eqref{eq:Dirac1} can be rewritten in the form of the Dirac equation in $1+1$ dimensions 
 \be
 \left ( i \gamma^0 \frac{\partial}{\partial x^0} +  i \gamma^1 \frac{\partial}{\partial x^1} - m \right) \psi = 0.
 \ee
In terms of Pauli matrices, the gamma matrices are given by
 \bea
 \gamma^0 &=& \sigma_3 \\
  \gamma^1 &=& i \sigma_1 
 \eea
which satisfy the usual anti-commutation relations:
 \bea
 \gamma^0 \gamma^1 +   \gamma^1 \gamma^0 &=&  0 \\
 \gamma^0 \gamma^0  =  - \gamma^1 \gamma^1&=&  I_2.\notag
 \eea
\subsection{Four types of neurons on three-dimensional lattice} 

A more physically relevant example corresponds to $K=4$ types of neurons on $D=3$ dimensional lattice with 
\bea
(W-I)^{a, {\bf j}}_{b, {\bf i}}  &=&  m {\cal X}^{a}_{0,b} \delta^{{\bf j}}_{{\bf i}}  \label{eq:weights}\\ 
{B}^{a, {\bf j}}_{b, {\bf i}}  &=&  \sum_{\mu=1}^3 {\cal X}^{a}_{\mu,b} \left (\delta^{{\bf j} -{\bf e}_\mu}_{\bf i}  - \delta^{{\bf j}}_{\bf i}  \right )      \notag
\eea
where the lattice vectors ${\bf e}_1=(1,0,0)$, ${\bf e}_2=(0,1,0)$, ${\bf  e}_3=(0,0,1)$). We also assume that the anti-symmetric tensors satisfy anti-commutation and commutation relations 
\bea
{\cal X}_{0} {\cal X}_{0} +{\cal X}_{0} {\cal X}_{0} &=& - 2 I_4 \notag\\
{\cal X}_{0}{\cal X}_{\mu}  - {\cal X}_{\mu}{\cal X}_{0} &=& 0 \label{eq:acomm}\\
{\cal X}_{\mu}{\cal X}_{\nu}  + {\cal X}_{\nu}{\cal X}_{\mu} &=& - 2 \delta_{\mu\nu} I_4 \notag
\eea
where $ \mu,\nu \in \{1,2,3\}$. Substituting \eqref{eq:weights} into \eqref{eq:eom4} yields 
\bea
\phi^{a}_{\bf{i}}(t+1) - \phi^{a}_{\bf{i}}(t)  &=&\sum_{\mu=1}^3  {\cal X}^a_{\mu,b} \left (\pi^{b}_{\bf{i}+{\bf e}_\mu}(t)  - \pi^{b}_{\bf{j}}(t)   \right )  + m {\cal X}^a_{0,b} \phi^{b}_{\bf{i}}(t)\label{eq:eom5}\\
\pi^{a}_{\bf{i}}(t+1) - \pi^{a}_{\bf{i}}(t) &=& - \sum_{\mu=1}^3 {\cal X}^a_{\mu,b} \left (   \phi^{b}_{\bf{i}}(t) -  \phi^{b}_{\bf{i}- {\bf e}_\mu}(t)\right )    -  m {\cal X}^a_{0,b} \pi^{b}_{\bf{i}}(t),\notag
\eea
and, in the continuous limit,
\bea
\frac{\partial \phi^a}{\partial x_0}  &=&\sum_{\mu=1}^3  {\cal X}^a_{\mu,b} \frac{\partial \pi^b}{\partial x_\mu}   + m {\cal X}^a_{0,b} \phi^{b} \label{eq:eom6}\\
\frac{\partial \pi^a}{\partial x_0} &=& - \sum_{\mu=1}^3 {\cal X}^a_{\mu,b} \frac{\partial \phi^b}{\partial x_\mu}     -  m {\cal X}^a_{0,b} \pi^{b}. \notag
\eea

The two real four-component vectors, $\phi^a$ and $\pi^a$, can be combined into two complex two-component vectors (or two-component spinors) by defining 
\bea
\tilde\phi \equiv \begin{pmatrix}
 {\phi}^{1} &+& i {\phi}^{2}\\
 {\phi}^{3} &+& i {\phi}^{4}\\
 \end{pmatrix} \label{eq:2spinors}\\
  \tilde\pi \equiv \begin{pmatrix}
 {\pi}^{1} &+& i {\pi}^{2}\\
 {\pi}^{3} &+& i {\pi}^{4}\\
 \end{pmatrix} \notag
\eea
and rewriting \eqref{eq:eom6} as 
\bea
\frac{\partial \tilde\phi^a}{\partial x_0}  &=&\sum_{\mu=1}^3  \tilde{\cal X}^a_{\mu,b} \frac{\partial \tilde\pi^b}{\partial x_\mu}   + m \tilde{\cal X}^a_{0,b} \tilde\phi^{b} \label{eq:eom7}\\
-\frac{\partial \tilde\pi^a}{\partial x_0} &=& \sum_{\mu=1}^3 \tilde{\cal X}^a_{\mu,b} \frac{\partial \tilde\phi^b}{\partial x_\mu}     + m \tilde{\cal X}^a_{0,b} \tilde\pi^{b} \notag
\eea
where $a,b \in \{1,2\}$, and $\tilde{\cal X}^a_{\mu}$ are four complex $2\times 2$ matrices that satisfy the same relations  \eqref{eq:acomm}. Furthermore, if we define a four-component spinor as 
\be
 \psi \equiv \begin{pmatrix}  \tilde{\phi}\\
 \tilde{\pi} \\
  \end{pmatrix} = \begin{pmatrix}
 {\phi}^{1} &+& i {\phi}^{2}\\
 {\phi}^{3} &+& i {\phi}^{4}\\
 {\pi}^{1} &+& i {\pi}^{2}\\
 {\pi}^{3} &+& i {\pi}^{4} 
 \end{pmatrix}.\label{eq:4spinor}
 \ee
 and 
\bea
\gamma^\mu &=& \sigma_1 \otimes \tilde{\cal X}_{\mu} \label{eq:gamma}\\
\gamma^0 &=& \sigma_3 \otimes I_2,\notag
\eea
where $\mu \in \{1,2,3\}$, then the two equations in \eqref{eq:eom7} can be combined
\be
\eta_{\mu\nu}\gamma^{\mu, a}_b \frac{\partial \psi^b}{\partial x_\nu}  -  m (I_2 \otimes \tilde{\cal X}_0)^a_{b} \psi^{b} = 0\label{eq:eom8}
\ee
where $\eta_{\mu\nu}$ is the flat metric with signature $\left(+,-, -, -\right)$ and $\mu,\nu \in \{0,1,2,3\}$. Due to \eqref{eq:acomm}, the gamma matrices \eqref{eq:gamma} satisfy the anti-commutation relation
 \be
 \gamma^\mu \gamma^\nu + \gamma^\nu \gamma^\mu = 2 \eta^{\mu\nu}. \label{eq:anticom}
\ee

For example, if the complex $2\times2$ matrices are 
\bea
\tilde{\cal X}_{\mu} &=& i \sigma_\mu\\
\tilde{\cal X}_0 &=& -i I_2,\notag
\eea
where $\mu \in \{1,2,3\}$, or equivalently, the real $4\times4$ matrices are
\bea
{\cal X}_{0} &=& I_2 \otimes i\sigma_2\notag\\ 
{\cal X}_{1}  &=& -  \sigma_1 \otimes i\sigma_2\\
{\cal X}_{2}   &=&  i\sigma_2 \otimes I_2 \notag\\
{\cal X}_{3} &=&- \sigma_3 \otimes  i\sigma_2,\notag
\eea
then the complex $4\times4$ matrices are
 \bea
 \gamma^0 &=&  \sigma_3  \otimes I_2 \notag\\
 \gamma^\mu &=&  \sigma_1 \otimes   \tilde{\cal X}_{\mu}  =  \sigma_1  \otimes i\sigma_\mu  
 \eea
and \eqref{eq:eom8} takes a familiar form of the Dirac equation in $3+1$ dimensions:
 \be
i \eta_{\mu\nu}\gamma^{\mu, a}_b \frac{\partial \psi^b}{\partial x_\nu}  -  m \psi^{a} = 0. \label{eq:Dirac}
 \ee

In the dual learning description of the Dirac field, the following linear combination of the trainable variables $\pi$ plays the role of biases:
\be
\beta^{a}_{\bf{i}} =  \sum_{\mu=1}^3 {\cal X}^a_{\mu,b} \left ( \pi^{b}_{{\bf i}+{\bf e}_\mu}  - \pi^{b}_{{\bf i}} \right ).    \label{eq:biases} 
 \ee
Due to the commutation relation \eqref{eq:acomm}, the operators $W$ and $B$ commute, and the equations of motion \eqref{eq:eom4} become:
 \bea
\phi^{a}_{\bf{i}}(t+1) - \phi^{a}_{\bf{i}}(t)   &=&m {\cal X}^a_{0,b}  \phi^{b}+\beta^{a} \\
\beta^{a}_{\bf{i}}(t+1) - \beta^{a}_{\bf{i}}(t)    &=& \left ( B B^T\right )^a_b \phi^{b} - m {\cal X}^a_{0,b} \beta^{b}.   \notag
\eea
By combining the two equations, in the continuous limit we obtain four uncoupled Klein-Gordon equations,
\be
\eta_{\mu\nu}\frac{\partial }{\partial x_\mu}\frac{\partial }{\partial x_\nu}  \phi^a({x}) +  m^2 \phi^a({x}) = 0,
\ee
while couplings takes places at the level of the $\phi$ and $\pi$ fields in the Dirac equation \eqref{eq:Dirac}.

 \section{Gauge field}\label{Sec:Gauge}

The discrete equations for activation and learning dynamics \eqref{eq:eom5} are invariant under the following transformation:
 \bea
 \phi^{a}_{\bf{i}}  \rightarrow  \tilde{\phi}^{a}_{\bf{i}}  = \left( {\cal X}^{\theta}_{0} \right )^a_b  \phi^{b}_{\bf{i}} \label{eq:global}\\
 \pi^{a}_{\bf{i}}  \rightarrow  \tilde{\pi}^{a}_{\bf{i}}  =  \left( {\cal X}^{\theta}_{0} \right )^a_b   \pi^{b}_{\bf{i}} \notag
 \eea
 where ${\cal X}^{\theta}_{0}$ is the matrix ${\cal X}_{0}$ raised to the power $\theta$. The invariance can be verified by multiplying \eqref{eq:eom5} by ${\cal X}^{\theta}_{0}$ and using the commutation relation \eqref{eq:acomm}. In the context of the Dirac field, \eqref{eq:global} describes a global $U(1)$ symmetry with phase angle represented by $ \theta \pi/2$. 
 
 The next step is to promote \eqref{eq:global} to a local transformation:
 \bea
 \phi^{a}_{\bf{i}}  \rightarrow  \tilde{\phi}^{a}_{\bf{i}}  =  \left( {\cal X}^{\theta_{\bf i}(t)}_{0} \right )^a_b   \phi^{b}_{\bf{i}} \label{eq:local}\\
 \pi^{a}_{\bf{i}}  \rightarrow  \tilde{\pi}^{a}_{\bf{i}}  = \left( {\cal X}^{\theta_{\bf i}(t)}_{0} \right )^a_b   \pi^{b}_{\bf{i}} \notag
 \eea
 where $\theta_{\bf i}(t)$ depends on both position ${\bf i}$ and time $t$. By multiplying \eqref{eq:eom5} by ${\cal X}^{\theta_{\bf i}(t)}_{0}$ and using the commutation relation \eqref{eq:acomm},  we find:
 \bea
\left( {\cal X}^{\theta_{\bf i}(t)- \theta_{\bf i}(t+1)}_{0} \right )^a_b \tilde{\phi}^{b}_{\bf{i}}(t+1) - \tilde{\phi}^{a}_{\bf{i}}(t)  &=&\sum_{\mu=1}^3  {\cal X}^a_{\mu,b} \left (\left( {\cal X}^{\theta_{\bf i}(t)- \theta_{\bf i + e_\mu}(t)}_{0} \right )^b_c  \tilde{\pi}^{c}_{\bf{i}+{\bf e}_\mu}(t)  - \tilde{\pi}^{b}_{\bf{i}}(t)   \right )  + m {\cal X}^a_{0,b} \tilde{\phi}^{b}_{\bf{i}}(t)\notag\\
\left( {\cal X}^{\theta_{\bf i}(t)- \theta_{\bf i}(t+1)}_{0} \right )^a_b \tilde{\pi}^{b}_{\bf{i}}(t+1) - \tilde{\pi}^{a}_{\bf{i}}(t) &=& - \sum_{\mu=1}^3 {\cal X}^a_{\mu,b} \left (   \tilde{\phi}^{b}_{\bf{i}}(t) -  \left( {\cal X}^{\theta_{\bf i}(t)- \theta_{\bf i - e_\mu}(t)}_{0} \right )^b_c \tilde{\phi}^{c}_{\bf{i}- {\bf e}_\mu}(t)\right )    -  m {\cal X}^a_{0,b} \tilde{\pi}^{b}_{\bf{i}}(t). \notag
\eea
Evidently, to ensure that the dynamical equations are invariant under local transformations \eqref{eq:local}, discrete derivatives in both space and time must also transform accordingly, i.e. a ``covariant'' derivative must be introduced. This can be achieved by introducing a gauge field that transforms as:
\bea
{A}_{0,{\bf i},b}^{a,{\bf j}}(t) &\rightarrow&  \tilde{A}_{0,{\bf i},b}^{a,{\bf j}}(t) =  {A}_{0,{\bf i},b}^{a,{\bf j}}(t)   + \left ( I - {\cal X}^{\theta_{\bf i}(t-1)- \theta_{\bf i}(t)}_{0}\right )_b^a \delta_{\bf i}^{\bf j}\\
{A}_{\mu,{\bf i},b}^{a,{\bf j}}(t) &\rightarrow&  \tilde{A}_{\mu,{\bf i},b}^{a,{\bf j}}(t) =  {A}_{\mu,{\bf i},b}^{a,{\bf j}}(t)    + \left ( I - {\cal X}^{\theta_{\bf i - e_\mu}(t)- \theta_{\bf i}(t)}_{0}\right )_b^a \delta_{\bf i}^{\bf j}
\eea
where $\mu \in \{1, 2, 3\}$. 

The resulting covariant equations then take the following form 
 \bea
{\phi}^{a}_{\bf{i}}(t\!+\!1)\!-\!{\phi}^{a}_{\bf{i}}(t)\!+\! {A}_{0,{\bf i},b}^{a,{\bf j}} (t) {\phi}^{b}_{\bf{j}}(t)  &\approx&\sum_{\mu=1}^3  {\cal X}^a_{\mu,b} \left ( {\pi}^{b}_{\bf{i}+{\bf e}_\mu}(t)\!-\!{\pi}^{b}_{\bf{i}}(t)\!+\!{A}_{\mu,{\bf i},c}^{b,{\bf j}} (t) {\pi}^{c}_{\bf j}(t)  \right )\!+\!m {\cal X}^a_{0,b} {\phi}^{b}_{\bf{i}}(t)\;\;\;\;\;\;\;\;\;\;\;\\
{\pi}^{b}_{\bf{i}}(t\!+\!1)\!-\!{\pi}^{a}_{\bf{i}}(t)\!+\! {A}_{0,{\bf i},b}^{a,{\bf j}} (t) {\pi}^{b}_{\bf{j}}(t)&\approx&\!-\!\sum_{\mu=1}^3 {\cal X}^a_{\mu,b} \left (   {\phi}^{b}_{\bf{i}}(t)\!-\!{\phi}^{b}_{\bf{i}- {\bf e}_\mu}(t)\!-\!{A}_{\mu,{\bf i},c}^{b,{\bf j}} (t) {\phi}^{c}_{\bf j}(t) \right )\!-\!m {\cal X}^a_{0,b} {\pi}^{b}_{\bf{i}}(t)\;\;\;\;\;\;\;\;\;\;\;\notag
\eea
under the assumption that fields vary continuously enough for discrete derivatives at nearby lattice points to be approximately equal. This introduces an additional term to the matrix and bias tensors:
\bea
(W-I)^{a, {\bf j}}_{{\bf i},b}  &=&  m {\cal X}^{a}_{0,b} \delta^{{\bf j}}_{{\bf i}} + {A}_{0,{\bf i},b}^{a,{\bf j}}  \label{eq:weights2}\\ 
{B}^{a, {\bf j}}_{{\bf i},b}  &=&  \sum_{\mu=1}^3 {\cal X}^{a}_{\mu,b} \left (\delta^{{\bf j} -{\bf e}_\mu}_{\bf i}  - \delta^{{\bf j}}_{\bf i} +{A}_{\mu,{\bf i},b}^{a,{\bf j}}  \right ).     \notag
\eea
From the learning side of the duality, this appears to be a natural generalization that makes the previously fixed weight 
$W$ and bias $B$ tensors dynamical.

For $U(1)$ symmetry we can take the weight matrix to be given by
\be
 (W-I)^{a, {\bf j}}_{{\bf i},b}  =   m {\cal X}^{a}_{0,b} \delta^{{\bf j}}_{{\bf i}} + {A}_{0,{\bf i},b}^{a,{\bf j}}  = m {\cal X}^{a}_{0,b} \delta^{{\bf j}}_{{\bf i}}  + {\cal X}^{a}_{0,b} {\cal A}_{0,{\bf i}}^{{\bf j}} 
 \ee
where ${\cal A}_{\mu,{\bf i}}^{{\bf j}}$  are diagonal for each $\mu \in \{0, 1, 2, 3\}$, the biases
\be
\beta^{a}_{\bf{i}} = {B}^{a, {\bf j}}_{{\bf i},b}  \pi^{b}_{\bf{j}}  =   \sum_{\mu=1}^3 {\cal X}^a_{\mu,b} \left ( \pi^{b}_{{\bf i}+{\bf e}_\mu}  - \pi^{b}_{{\bf i}} + {\cal X}^{a}_{0,b} {\cal A}_{\mu,{\bf i}}^{{\bf j}}  \pi^{b}_{\bf{j}} \right ).    \label{eq:biases2} 
 \ee
 and then the loss functions are given by
\be
H_a(\phi,\pi) =  \frac{1}{2} \phi_a^{\bf i} \phi^a_{\bf i} -   \pi_a^{\bf i} {\cal X}^a_{0,b} \left ( m   +  {\cal A}_{0,{\bf i}}^{{\bf i}}\right )\pi^b_{\bf i} \label{eq:losses2}
\ee
instead of \eqref{eq:losses} and the corresponding Dirac equation is
 \be
 i \eta_{\mu\nu}\gamma^{\mu, a}_b \left ( \frac{\partial}{\partial x_\nu}+ i {\cal A}_{\mu}\right )  \psi^b  -  m \psi^{a} = 0. \label{eq:Dirac2}
 \ee
Note that ${\cal A}_{\mu}$ plays the role of a background gauge field, whose dynamical equations must be introduced independently, perhaps, at the level of the learning dynamics.

\section{Discussion}\label{Sec:Discussion}

In this article, we established a duality relation between classical systems and learning systems, introducing a novel framework to understand the connection between physics and learning. Specifically, we demonstrated that the Hamilton's equations governing the dynamics of position and momentum variables correspond to equations governing the activation and learning dynamics of non-trainable and trainable variables, respectively. The duality is not merely formal; it can also serve as a useful tool for modeling, for example, field theories through the activation dynamics of the state of neurons and the learning dynamics of biases in a neural network. Furthermore, this approach offers a conceptual bridge between physical and learning systems, providing a fresh perspective on both domains.

In particular, the duality was applied to model both the Klein-Gordon and Dirac equations. On the physics side, these equations describe the dynamics of a scalar field and a four-component spinor, respectively. On the learning side, they describe the dynamics of the state of neurons and biases, using symmetric (for Klein-Gordon) and anti-symmetric (for Dirac) factors of a fixed weight tensor. In the case of gauge fields, the weight and bias tensors become dynamical, offering insights into how gauge field theories might emerge from the learning dynamics of neural networks.

It is important to emphasize that the theories considered in this article are entirely classical and do not incorporate any quantum or gravitational effects. However, this raises an intriguing question: could quantum mechanics and/or general relativity emerge from the activation and learning dynamics of neural networks under certain conditions, or perhaps in some limit?  More generally, could all physical and perhaps even non-physical phenomena be viewed as emergent phenomena arising from the learning dynamics of neural networks?  This line of inquiry is not about training a neural network to directly solve dynamical equations or to mimic the behavior of physical or non-physical systems; rather, it is about understanding whether these equations or behaviors can emerge, in some limit, from the activation and learning dynamics inherent to neural networks.

These questions have been partially addressed with respect to physical phenomena such as quantum mechanics \cite{WaNN, Quantumness, TTQG}, general relativity \cite{WaNN, TTQG}, and self-organized criticality \cite{Westerhout, Kukleva}, as well as biological phenomena like biological evolution \cite{Koonin}, the origin of life \cite{Koonin2}, programmed death and replication \cite{Grabovsky} and phase transitions \cite{Romanenko}. Collectively, these results, along with the duality developed in this article, suggest that there may be a much deeper connection between neural networks and fundamental physics, with artificial neural networks providing a framework for modeling and understanding physical systems, and physical theories offering a framework for modeling and understanding learning systems. For example, the usefulness of physical symmetries has recently been shown to significantly improve the behavior of autonomous vehicles in complex environments \cite{Andrejic}, but it remains to be seen whether the opposite is also true --- that complex physical systems can be effectively modeled as learning systems \cite{Gusev}.

There are several related approaches that explore the idea of emergent field theories \cite{Wilson, Susskind, Kogut, Creutz}, emergent quantum mechanics \cite{Adler, tHooft, Caticha, Quantum}, and emergent gravity \cite{Jacobson, Padmanabhan, Verlinde, Gravity}. These approaches have opened up new avenues for studying the nature of physical reality, suggesting that what we perceive as fundamental fields may emerge from more fundamental and perhaps even discrete structures such as particles, strings, or even neurons. In this context, the current paper can be viewed as a continuation of these previous works, with a specific emphasis on the emergence of scalar, spinor and gauge fields from the learning dynamics of neural networks, whether they are artificial, biological, or fundamental.

{\it Acknowledgements.} The author is grateful to Yaroslav Gusev, Ekaterina Kukleva, and Mikhail Katsnelson for their invaluable insights and constructive discussions that greatly contributed to this work.

\end{document}